\documentclass[aps,prb,twocolumn,superscriptaddress,showpacs]{revtex4-1}

\usepackage{graphicx}
\usepackage{calc}
\usepackage{bm}
\usepackage{color}

\begin{document}

\title{Multivalued current-phase relationship in a.c. Josephson effect for a three-dimensional Weyl semimetal WTe$_2$.}

\author{O.O.~Shvetsov}
\affiliation{Institute of Solid State Physics of the Russian Academy of Sciences, Chernogolovka, Moscow District, 2 Academician Ossipyan str., 142432 Russia}
\affiliation{Moscow Institute of Physics and Technology, Institutsky per. 9, Dolgoprudny, 141700 Russia}
\author{A.~Kononov}
\author{A.V.~Timonina}
\author{N.N.~Kolesnikov}
\author{E.V.~Deviatov}
\affiliation{Institute of Solid State Physics of the Russian Academy of Sciences, Chernogolovka, Moscow District, 2 Academician Ossipyan str., 142432 Russia}

\date{\today}

\begin{abstract}
We  experimentally study electron transport between two superconducting indium leads, coupled to a single WTe$_2$ crystal, which is  a three-dimensional  Weyl semimetal. We demonstrate Josephson current in  long 5~$\mu$m In-WTe$_2$-In junctions, as confirmed by
the observation of integer (1,2,3) and fractional (1/3, 1/2, 2/3) Shapiro steps under microwave irradiation. Demonstration of fractional a.c. Josephson effect indicates multivalued character of the current-phase relationship, which we connect with Weyl topological surface states contribution to Josephson current. In contrast to topological insulators and Dirac semimetals,  we do not  observe $4\pi$ periodicity in a.c. Josephson effect for WTe$_2$  at different frequencies and power, which might reflect chiral character of the Fermi arc surface states in Weyl semimetal.
\end{abstract}


\maketitle

\section{Introduction}

Like other topological materials~\cite{hasan,zhang,das,chiu}, Weyl semimetals are characterized by topologically protected  surface states. These states originate as  Fermi arcs, which  connect the projections of Weyl points on the surface Brillouin zone in k-space~\cite{armitage}. In contrast to helical surface states in topological insulators~\cite{hasan}, Weyl states  inherit the chiral property of the Chern insulator edge states~\cite{armitage}.  Fermi arcs have been experimentally demonstrated by angle-resolved photoemission spectroscopy, e.g. for MoTe$_2$ and WTe$_2$ three-dimensional crystals~\cite{wang,wu}. 

The concept of Fermi arcs  has been used to explain the magnetotransport experiments~\cite{mazhar,wang-miao}. Unfortunately, Weyl and Dirac semimetals are conductors with gapless bulk excitations~\cite{armitage}, so it is a problem to reliably distinguish the bulk and surface transport properties. On the other hand, the edge current contribution can be retrieved even for systems with conducting bulk by analyzing Josephson current behavior~\cite{yakoby,kowen,inwte}. Edge  state transport is responsible for Josephson current in 1-2~$\mu$m long superconductor-normal-superconductor (SNS) junctions in graphene~\cite{calado,borzenets}.  For the Cd$_3$As$_2$ Dirac semimetal, observation  of $\pi$ and $4\pi$ periodic current-phase relationship has been reported  in Al-Cd$_3$As$_2$-Al and Nb-Cd$_3$As$_2$-Nb  junctions~\cite{rfpan,rfli}.  In this case, the fractional a.c. Josephson effect ($\pi$ periodicity) is connected with interference between the bulk and surface supercurrent contributions, while the disappearance of $N=1$ Shapiro step ($4\pi$ periodicity) reflects the the helical nature of topological surface states in Dirac semimetals~\cite{rfpan,rfli}.  Thus, it seems to be reasonable to study a.c.  Josephson effect in SNS junctions, fabricated on a  Weyl semimetal surface.

Here, we experimentally study electron transport between two superconducting indium leads, coupled to a single WTe$_2$ crystal, which is  a three-dimensional  Weyl semimetal. We demonstrate Josephson current in  long 5~$\mu$m In-WTe$_2$-In junctions, as confirmed by
the observation of integer (1,2,3) and fractional (1/3, 1/2, 2/3) Shapiro steps under microwave irradiation. Demonstration of fractional a.c. Josephson effect indicates multivalued character of the current-phase relationship, which we connect with Weyl topological surface states contribution to Josephson current. In contrast to topological insulators and Dirac semimetals,  we do not  observe $4\pi$ periodicity in a.c. Josephson effect for WTe$_2$  at different frequencies and power, which might reflect chiral character of the Fermi arc surface states in Weyl semimetal.

\section{Samples and technique}

\begin{figure}
\includegraphics[width=\columnwidth]{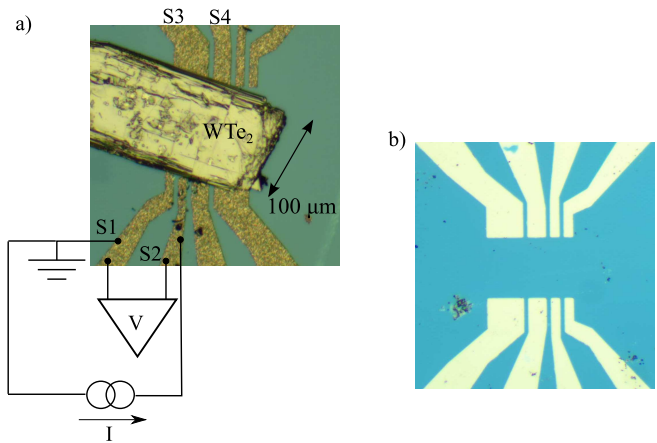}
\caption{(Color online) (a) Top-view image of the sample with sketch of electrical connections. 10~$\mu$m wide indium superconducting leads  (S1-S4) are separated by 5~$\mu$m intervals on the insulating SiO$_2$ substrate.  In-WTe$_2$-In junctions are fabricated on the bottom surface of a WTe$_2$ crystal  by weak pressing a crystal ($\approx 0.5\mbox{mm}\times 100\mu\mbox{m} \times 0.5 \mu\mbox{m} $) to the indium leads pattern.  Charge transport is investigated between two superconducting electrodes in a four-point technique: the S1 electrode  is grounded; a current $I$ is fed through the S2; a voltage drop $V$ is measured between these S1 and S2 electrodes by independent wires because of low normal In-WTe$_2$-In resistance.  (b) Image of  the leads pattern without a WTe$_2$ crystal.  
}
\label{sample}
\end{figure}

WTe$_2$ compound was synthesized from elements by reaction of metal with tellurium vapor in the sealed silica ampule. The WTe$_2$ crystals were grown by the two-stage iodine transport~\cite{growth1}, that previously was successfully applied~\cite{growth1,growth2} for growth of other metal chalcogenides like NbS$_2$ and CrNb$_3$S$_6$. The WTe$_2$ composition is verified by energy-dispersive X-ray spectroscopy. The X-ray diffraction (Oxford diffraction Gemini-A, MoK$\alpha$) confirms $Pmn2_1$ orthorhombic single crystal WTe$_2$ with lattice parameters $a=3.48750(10)$~\AA, $b= 6.2672(2)$~\AA, and $c=14.0629(6)$~\AA. We check by standard magnetoresistance measurements that our WTe$_2$ samples demonstrate large, non-saturating positive magnetoresistance $\rho(B)-\rho(B=0)/\rho(B=0)$  in normal magnetic field, which goes to zero in parallel one,  as it has been shown for WTe$_2$ Weyl semimetal~\cite{mazhar}, see Ref.~\cite{ndwte} for details of magnetoresistance measurements.

A sample sketch is presented in Fig.~\ref{sample}. Superconducting leads are formed by lift-off technique after thermal evaporation of 100~nm indium on the insulating SiO$_2$ substrate, see Fig.~\ref{sample} (b). A WTe$_2$ single crystal ($\approx 0.5 \mbox{mm}\times 100\mu\mbox{m} \times 0.5 \mu\mbox{m} $ dimensions) is  placed on the indium leads pattern, and is weakly pressed by another Si/SiO$_2$ substrate. The substrates are kept  strictly parallel by external metallic frame to avoid sliding of the WTe$_2$ crystal, which is verified in optical microscope. As a result,  planar In-WTe$_2$ junctions  are formed at the bottom surface of the crystal WTe$_2$, being separated by 5~$\mu$m intervals, as depicted in Fig.~\ref{sample}. 

The obtained  In-WTe$_2$-In  SNS structures should be regarded as long $\xi<< L$ diffusive $L>l_e$ ones: the $L$ value exceeds the mean free path~\cite{wte2mobility} in WTe$_2$ $l_e\approx$1~$\mu$m, so it should be compared~\cite{kulik-long,dubos} with the coherence length of the diffusive SNS junction $\xi=(l_e \times \hbar v_F^N/\pi\Delta_{in})^{1/2}\approx 200$~nm, where Fermi velocity is  $v_F^N\approx 1.5\cdot10^7 \frac{cm}{s}$ from ARPES data~\cite{bruno}, and $\Delta_{In}=0.5$~meV is the indium superconducting gap~\cite{indium}.  This estimation is even stronger for smaller $l_e$, which can be expected~\cite{ali-mag-res} from the magnetoresistance behavior of our samples~\cite{inwte,ndwte}.

Charge transport is investigated between two superconducting indium leads in a four-point technique. An example of electrical connections is presented in Fig.~\ref{sample} (a): the S1 electrode  is grounded; a current $I$ is fed through the S2; a voltage drop $V$ is measured between these S1 and S2 electrodes by independent wires. In this connection scheme, all  the wire resistances are excluded, which is necessary for low-impedance  In-WTe$_2$-In junctions (below 0.5 Ohm normal resistance in the present experiment).  The measurements are performed in standard He$^4$ cryostat in the temperature range 1.4~K -- 4.2~K. The indium leads are superconducting below the  critical temperature~\cite{indium} $T_c\approx 3.4~K$.

\section{Experimental results}

\begin{figure}
\includegraphics[width=\columnwidth]{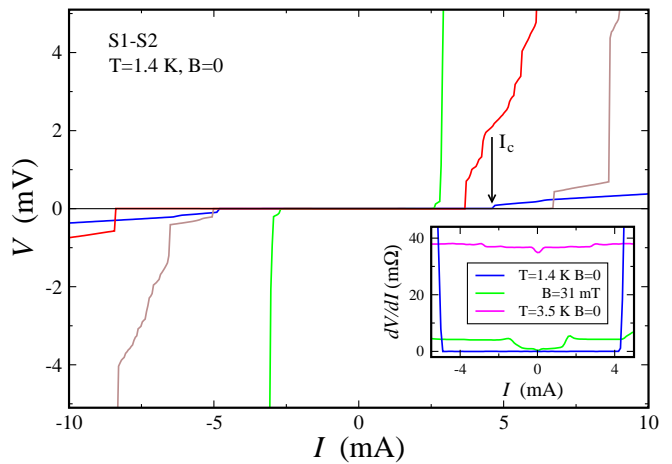}
\caption{(Color online)  Examples of  $I-V$ characteristics for different samples,  obtained for 5~$\mu$m long In-WTe$_2$-In junction between the superconducting leads S1 and S2, as depicted in Fig.~\protect\ref{sample}.  A clear Josephson behavior can be seen in zero magnetic field at 1.4~K$<T_c$: there is no  resistance at low  currents, it appears above $ \pm I_c\approx 2-8$~mA for different samples.  The jump positions are subjected to small hysteresis with the sweep direction, so they are slightly different for two $I-V$ branches. Inset: $dV/dI(I)$ characteristics for the S1-WTe$_2$-S2 junction at minimal $T=1.4$~K$<T_c$ (the blue curve) and at $T=$3.5~K$>T_c$ (the red curve) in zero magnetic field, and at the critical field $B=31$~mT at minimal  $T=1.4$~K (the green one).
} 
\label{IV}
\end{figure}

\subsection{$I-V$ curves}

To obtain $I-V$ characteristics,  we sweep the dc current $I$ and measure the voltage drop $V$. Fig.~\ref{IV} presents $I-V$ examples  for different samples in zero magnetic field and at low temperature  1.4~K$<T_c$. 

The curves in Fig.~\ref{IV} clearly demonstrate Josephson effect in   unprecedentedly long  $L=$5~$\mu$m In-WTe$_2$-In  junctions: (i) by the four-point connection scheme we directly demonstrate zero resistance  region at low  currents (with $\pm 0.05$~m$\Omega$ accuracy, see the inset); (ii) the non-zero resistance appears as sharp jumps  at current values $ \pm I_c\approx 2-8$~mA for different samples; (iii) $I-V$ curve can be switched to standard Ohmic behavior, if   superconductivity is suppressed by temperature  or magnetic field, as it is demonstrated in the inset to Fig.~\ref{IV}.

The obtained $I_c$ values differ within 10\% in different coolings for a given sample. They are much smaller than the critical current for the indium leads, which can be estimated as  $\approx 30$~mA  for the leads' dimensions and the known~\cite{in-current} maximum value $j\approx 3\times 10^6$A/cm$^2$ for indium. There are also  small jumps in the resistive state at intermediate currents $I_c<I<30$~mA for some samples, see Fig.~\ref{IV}.

\subsection{$I-V$ curves under microwave irradiation}

 \begin{figure}
\includegraphics[width=\columnwidth]{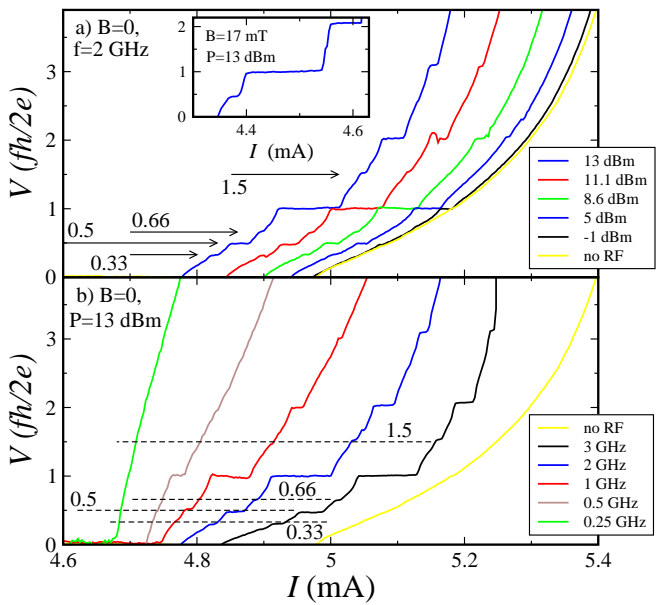}
\caption{(Color online) A.c. Josephson effect in an  In-WTe$_2$-In SNS junction at minimal 1.4~K temperature. Shapiro steps appear at $V=N hf/2e$. Integer steps at $N=1,2,3,..$  are typical for SNS junctions, while the fractional $N=3/2, 2/3, 1/2, 1/3$ ones indicate a multivalued nonsinusoidal character of the current-phase relationship~\cite{rfpan,rfli,bezryadin}. (a) At fixed frequency 2 GHz,  the $N=1$ step appears first. At higher power, $N=2,3$ ones appear together with the fractional $N=1/2$ step. The fractions $N=3/2, 2/3,1/3$ can only be seen at maximum power, and they can be suppressed by lowest 17~mT magnetic field, as demonstrated in the inset. (b) The $N=1$  is the most robust also when decreasing the frequency at fixed power 13 dBm. The curves are shifted for clarity in (b).
} 
\label{rf}
\end{figure}

The main experimental finding is the observation of fractional a.c.  Josephson effect, as it is depicted in Fig.~\ref{rf}. The sample is illuminated by microwave (rf) radiation through an open coaxial line. For the fixed  frequency,  see Fig.~\ref{rf} (a), rising of the radiation power shifts $I_c$ to lower currents. Simultaneously,  Shapiro steps appears, which are placed at  $V=N hf/2e$, as it should be expected for typical SNS junctions with trivial $2\pi$ periodicity in current-phase relationship $I_J=I_csin(\phi)$. 

In addition to the  steps at integer $N=1,2,3..$, we observe  half-integer $N=1/2,3/2$ ones, i.e., $\pi$ periodicity in a.c. Josephson effect. It usually appears due to interference effects~\cite{rfpan,rfli}, for example, for the double-slit geometries in superconducting quantum interference devices (SQUID)~\cite{bezryadin,iran}. The situation is even more complicated at high power: there are clear-visible $N=1/3,2/3$ steps, which obviously indicates a multivalued nonsinusoidal character of the current-phase relationship~\cite{bezryadin,iran}.

Fig.~\ref{rf} also demonstrates dependence of integer and fractional Shapiro steps on the microwave power (a), magnetic field (inset to (a)), and microwave frequency (b).  The  fractional $N=1/3,2/3$ steps are the weakest: they can be suppressed by lowest magnetic field, while the  $N=1/2$ one is as robust as the integer $N=2,3$ Shapiro steps, see the inset to Fig.~\ref{rf} (a). Also, while decreasing the frequency at constant power, $N=1/3,2/3$ steps disappear first, which is demonstrated in Fig.~\ref{rf} (b). On the other hand, the integer  $N=1$ step is the most stable: it appears at lowest power and frequency, see Fig.~\ref{rf}, and it is the strongest at highest ones. This robustness of the $N=1$ Shapiro step is just the opposite to the observed $4\pi$ periodicity ($N=1$ disappearance) in Al-Cd$_3$As$_2$-Al or Nb-Cd$_3$As$_2$-Nb  junctions~\cite{rfpan,rfli}.

\begin{figure}
\includegraphics[width=\columnwidth]{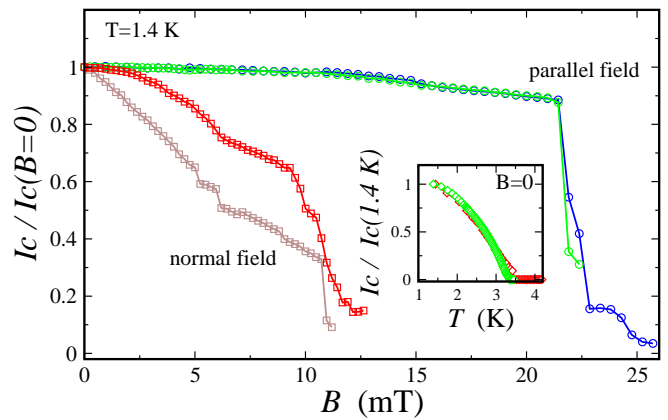}
\caption{(Color online) Dependencies  of the maximum supercurrent $I_c$  on the magnetic field for two different samples.  $I_c(B)$  pattern crucially depends on the magnetic field orientation to the  In-WTe$_2$-In junction plane: it is  strong for the perpendicular field, while it is very slow (within 10\% until the critical field) for the parallel orientation (magnetic field is parallel to the  $b$ axis of Wte$_2$, as depicted in Fig.~\protect\ref{discussion}). The curves are obtained at minimal 1.4~K temperature. All the experimental points are well reproducible, variation of $I_c$  is below the symbol size. Inset demonstrates the maximum supercurrent $I_c$ as function of temperature in zero magnetic field for  two different samples. $I_c (T)$ monotonously  falls to zero at 3.5~K, the $I_c (T)$ dependence is obviously slower than the linear function of $T$ even in the high-temperature limit $T\sim T_c$. 
} 
\label{Ic_BT}
\end{figure}

\subsection{Check of possible fabrication defects}

First of all,  we should experimentally exclude artificial reasons for the observed multivalued character of the current-phase relationship, particularly possible fabrication defects, like multiple indium shortings in the junction plane. The thickness of the indium film is chosen to be much smaller than the leads separation (100~nm$<< 5\mu$m) to avoid parasite shorting of In leads. Moreover, we do not see Josephson current in 5~$\mu$m long In-Cd$_3$As$_2$-In junctions, prepared in the same technique~\cite{cdasJ}, so the observed behavior is specific for WTe$_2$.

The crucial arguments can be obtained from the maximum supercurrent $I_c$ behavior with temperature or magnetic field. 

To analyze $I_c (B,T)$ behavior, we use $dV/dI(I)$ characteristics like ones presented in the right inset to Fig.~\ref{IV}: the dc current is additionally modulated by a low ac component (100~nA, 10~kHz), an ac  part of $V$ ($\sim dV/dI$) is detected by a lock-in amplifier. We have checked, that the lock-in signal is independent of the modulation frequency in the 6~kHz -- 30~kHz range, which is defined by applied ac filters.   To obtain $I_c$ values with high accuracy for given $(B,T)$ values,  we sweep current $I$ ten times from zero (superconducting state) to above $I_c$ (resistive state), and then determine  $I_c$ as the average value of $dV/dI$ jump positions in different sweeps. The results are presented in Fig.~\ref{Ic_BT}. All the experimental points are well reproducible, variation of $I_c$ in different sweeps is below the symbol size  for data in Fig.~\ref{Ic_BT}. 

(i) The experimental $I_c (T)$ in the inset to Fig.~\ref{Ic_BT} is inconsistent with indium shortings, because  $I_c (T)$ does not demonstrate strong decay in the high-temperature limit $T\sim T_c$, which is expected~\cite{kulik-long,dubos} for long diffusive SNS junctions. Instead, the experimental $I_c (T)$ dependence is even slower than the linear function of $T$ in Fig.~\ref{Ic_BT} (a). Similar behavior has been also demonstrated in long (1.5-2~$\mu$m) graphene SNS junctions~\cite{calado,borzenets}, where it has been attributed to topological edge state transport. 

(ii)  To our surprise, $I_c(B)$  pattern crucially depends on the magnetic field orientation to the  In-WTe$_2$-In junction plane, see Fig.~\ref{Ic_BT}. If the magnetic field is perpendicular to the plane, strong suppression of $I_c(B)$ is observed, as it can be expected for standard Josephson junctions due to the pair breaking effect~\cite{cuevas}.  In contrast,  $I_c(B)$ is diminishing very slowly (within 10\% until the  critical field)   for the parallel magnetic field, which  indicates interference effects, like in non-symmetric double-slit SQUID  geometries~\cite{yakoby,kowen}. It indicates, that the effective SQUID area is perpendicular to the junction plane and, thus,  it can not be formed by parasite In shortings: if several fabrication defects connected the leads on the SiO$_2$ surface, they could  form   SQUID-like geometry in the junction plane only.

\section{Discussion}

\begin{figure}
\centerline{\includegraphics[width=\columnwidth]{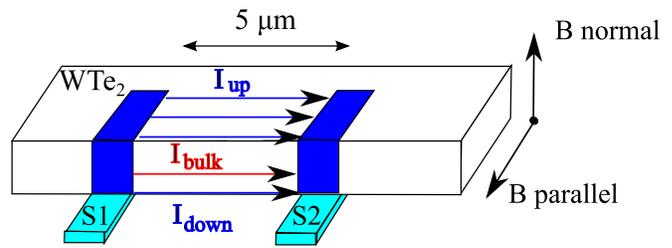}}
\caption{(Color online) Schematic diagram of surface state (blue arrows) and bulk (the red one) contributions to Josephson current, which are responsible for fractional a.c. Josephson effect at $N=1/3,2/3$ and 1/2 (see the main text). 
}
\label{discussion}
\end{figure}

Since we can exclude parasite shortings in the junction plane, we should connect the observed multivalued character of the current-phase relationship   with non-trivial distribution~\cite{yakoby,kowen,inwte,rfpan} of the Josephson current within the WTe$_2$ crystal, i.e., with topological surface states~\cite{armitage,wang,mazhar,wang-miao}.  

The interference ($\pi$ periodicity) can appear if both surface and bulk carriers transfer Josephson current in parallel, see  Fig.~\ref{discussion}, as it has been proposed for Cd$_3$As$_2$ Dirac semimetal~\cite{rfpan,rfli}. This picture gives qualitatively reasonable results: (i)  Parallel magnetic field induces a phase shift between surface and bulk channels for Josephson current, which leads to slow $I_c(B)$ damping, as we observe in Fig.~\ref{Ic_BT}. If the magnetic field is perpendicular to the WTe$_2$ crystal plane, both surface and bulk channels are in phase. (ii) Half-integer $N=1/2,3/2$ Shapiro steps appears in Fig.~\ref{rf} due to interference between bulk and surface channels, similarly to Cd$_3$As$_2$ Dirac semimetal~\cite{rfpan,rfli}.

However, the clear visible   Shapiro steps at fractional $N=1/3,2/3$ indicates that the interference scheme for Josephson current is more complicated~\cite{bezryadin,iran} for WTe$_2$. This is the reason to consider also the surface states on the opposite sample surface, see  Fig.~\ref{discussion}, which are not independent in Weyl semimetal~\cite{armitage}. Due to the necessity of the surface state coupling, the contribution of this channel can only be seen at maximum microwave power and frequency in Fig.~\ref{rf}.

Another evidence on the Weyl specifics of the WTe$_2$ surface states is the fact, that we do not  observe $4\pi$ periodicity in a.c. Josephson effect: the integer  $N=1$ Shapiro step is the strongest  one at highest power in Fig.~\ref{rf}, while the  maximum power value covers the range of $N=1$ disappearance in Al-Cd$_3$As$_2$-Al and Nb-Cd$_3$As$_2$-Nb  junctions~\cite{rfpan,rfli}. In the latter case, $4\pi$ periodicity is connected~\cite{rfpan,rfli} with the helical nature of topological surface states in Dirac semimetals, while Weyl surface states  inherit the chiral property of the Chern insulator edge states~\cite{armitage}.

 Because of topological protection,  Weyl surface states  can  efficiently transfer the Josephson current, which  appears in   slow $I_c(T)$ decay in the inset to Fig.~\ref{Ic_BT}. This is another argument for the surface states, since the bulk supercurrent contribution should demonstrate strong exponential decay~\cite{kulik-long,dubos} in the high-temperature limit $T\sim T_c$  for long diffusive SNS junctions.

\section{Conclusion}

As a conclusion, we experimentally study electron transport between two superconducting indium leads, coupled to a single WTe$_2$ crystal, which is  a three-dimensional  Weyl semimetal. We demonstrate Josephson current in  long 5~$\mu$m In-WTe$_2$-In junctions, as confirmed by
the observation of integer (1,2,3) and fractional (1/3, 1/2, 2/3) Shapiro steps under microwave irradiation. Demonstration of fractional a.c. Josephson effect indicates multivalued character of the current-phase relationship, which we connect with Weyl topological surface states contribution to Josephson current. In contrast to topological insulators and Dirac semimetals,  we do not  observe $4\pi$ periodicity in a.c. Josephson effect for WTe$_2$  at different frequencies and power, which might reflect chiral character of the Fermi arc surface states in Weyl semimetal.

\acknowledgments
We wish to thank Yu.S.~Barash, Ya.~Fominov, V.T.~Dolgopolov for fruitful discussions, and S.S~Khasanov for X-ray sample characterization.  We gratefully acknowledge financial support by the RFBR (project No.~16-02-00405) and RAS.

\end{document}